\documentclass[bibyear]{aa} 
\usepackage{amsmath}	
\usepackage{amssymb}	
\makeatletter
\renewcommand\@biblabel[1]{}
\makeatother
\usepackage{graphicx}
\usepackage{txfonts}
\begin{document}

   \title{Mirach's Goblin: Discovery of a dwarf spheroidal galaxy behind the Andromeda galaxy}


   \author{David Mart\'\i nez-Delgado\inst{1},
Eva K. Grebel\inst{1},
Behnam Javanmardi\inst{2}, Walter Boschin\inst{3,4.5}, Nicolas Longeard\inst{6}, Julio A. Carballo-Bello\inst{7}, Dmitry Makarov\inst{8}, Michael A. Beasley\inst{4,5},  Giuseppe Donatiello\inst{9}, Martha P. Haynes\inst{10}, Duncan A. Forbes\inst{11}, Aaron J.\ Romanowsky\inst{12,13}
          }

\institute{$^{1}$Astronomisches Rechen-Institut, Zentrum f\"ur Astronomie der Universit\"at Heidelberg,  M{\"o}nchhofstr. 12-14, 69120 Heidelberg, \\ ~~Germany\\
$^{2}$ School of Astronomy, Institute for Research in Fundamental Sciences (IPM), Tehran, 19395-5531, Iran\\
$^{3}$ Fundaci\'on G. Galilei - INAF (Telescopio Nazionale Galileo), Rambla J. A. Fern\'andez P\'erez 7, E-38712 Bre\~na Baja (La Palma), Spain\\
$^{4}$Instituto de Astrof\'isica de Canarias (IAC), Calle V\'ia L\'actea s/n, E-38205 La Laguna, Tenerife; Spain\\
$^{5}$Departamento de Astrof\'\i sica, Universidad de La Laguna (ULL), E-38206 La Laguna, Tenerife; Spain\\ 
$^{6}$ Universit\'e de Strasbourg, CNRS, Observatoire astronomique de Strasbourg, UMR 7550, F-67000 Strasbourg, France\\
$^{7}$Instituto de Astrof\'isica, Facultad de F\'isica, Pontificia Universidad Cat\'olica de Chile, 782-0436 Macul, Santiago, Chile\\ 
$^{8}$ Special Astrophysical Observatory, Nizhniy Arkhyz, Karachai-Cherkessia 369167, Russia\\
$^{9}$ Nuovo Orione, 72024 Oria, Italy \\
$^{10}$Cornell Center for Astrophysics and Planetary Science, Space Sciences Building, Cornell University, Ithaca, NY 14853
, USA\\
$^{11}$ Centre for Astrophysics \& Supercomputing, Swinburne University of Technology, Hawthorn VIC 3122, Australia\\
$^{12}$ Department of Physics \& Astronomy, San Jos\'e State University, One Washington Square, San Jose, CA 95192, USA\\
$^{13}$ University of California Observatories, 1156 High Street, Santa Cruz, CA 95064, USA\\}



   \date{}

 
  \abstract
   {It is of broad interest for galaxy formation theory to carry out a full inventory of the numbers and properties of dwarf galaxies in the Local Volume, both satellites and isolated ones.}
   {Ultra-deep imaging in wide areas of the sky with small amateur telescopes can help to complete the census of these hitherto unknown low surface brightness galaxies, which cannot be detected by the current resolved stellar population and HI surveys. We report the discovery of Donatiello~I, a dwarf spheroidal galaxy located one degree from the star Mirach ($\beta$ And) in a deep image taken with an amateur telescope.}
   {The color--magnitude diagram obtained from follow-up observations obtained with the Gran Telescopio Canarias (La Palma, Spain) reveals that this system is beyond the Local Group and is mainly composed of old stars. The absence of young stars and HI emission in the ALFALFA survey are typical of quenched dwarf galaxies. Our photometry suggests a distance modulus for this galaxy of $(m-M)=27.6 \pm 0.2$ (3.3 Mpc), although this distance  cannot yet be established securely owing to the crowding effects in our color--magnitude diagram. At this distance, Donatiello~I's absolute magnitude ($M_{V} =-8.3$), surface brightness 
($\mu_{V}=26.5$ mag arcsec$^{-2}$) and stellar content are similar to  the``classical" Milky Way companions Draco or Ursa Minor. }
   {The projected position and distance of Donatiello~I are  consistent with being a dwarf satellite of the closest S0-type galaxy NGC 404 (``Mirach's Ghost"). Alternatively, it could be one of the most isolated quenched dwarf galaxies reported so far behind the Andromeda galaxy. }
   {}

 \authorrunning{Mart\'\i nez-Delgado et al.}

  \maketitle
%

\section{Introduction}

The $\Lambda$-CDM paradigm predicts a large number of small dark matter halos in the Local Volume, but it is unclear how many of them are associated with luminous baryons in the form of stars.  It is thus of broad interest for galaxy formation theory to carry out a full inventory of the numbers and properties of dwarf galaxies, both satellites and isolated ones. The observations are certainly biased: the earlier surveys of the Local Group spirals based on
photographic plates were limited in surface brightness to about
25.5 mag arcsec$^{-2}$ (Whiting 2007). For the Milky Way (MW) and Andromeda (M31), the recent discoveries of dwarf satellites were
made using stellar density maps of resolved stars, counted in selected areas of the color-magnitude diagrams (CMDs) from large scale photometric survey data, such as the Sloan Digital Sky Survey (SDSS; Abazajian et al. 2009), the Panoramic Survey Telescope and Rapid Response System (Pan-STARRs; Chambers et al. 2017) and more recently the Dark Energy Survey (DES; Abbott et al. 2018) and the Pan-Andromeda Archaeological Survey (PAndAs) of the Andromeda galaxy with the wide-field imager on the Canada French Hawaii Telescope (CFHT) (McConnachie et al. 2009; Ibata et al. 2014; Martin et al. 2013). 

In the case of our Galaxy, the main tracers are A-type stars 2--3 magnitudes fainter than the main sequence (MS)-turnoff of the old stellar population,
which led to the discovery of the ultra-faint population of dwarf satellites and very diffuse halo stellar sub-structures (Newberg et al. 2002; Willman et al. 2005; Belokurov et al. 2006). However, the relatively shallow CMDs from the SDSS (with a $g$-band limiting magnitude for point sources of $\sim$ 21.5)  and the significant contamination of the MS-turnoff CMD locus by distant blue galaxies (e.g. see Fig. 1 in
Martinez-Delgado et al. 2004), make it difficult to complete the census of faint MW dwarf companions for distances larger than 100 kpc. 

\begin{figure}
\includegraphics[width=0.48\textwidth]{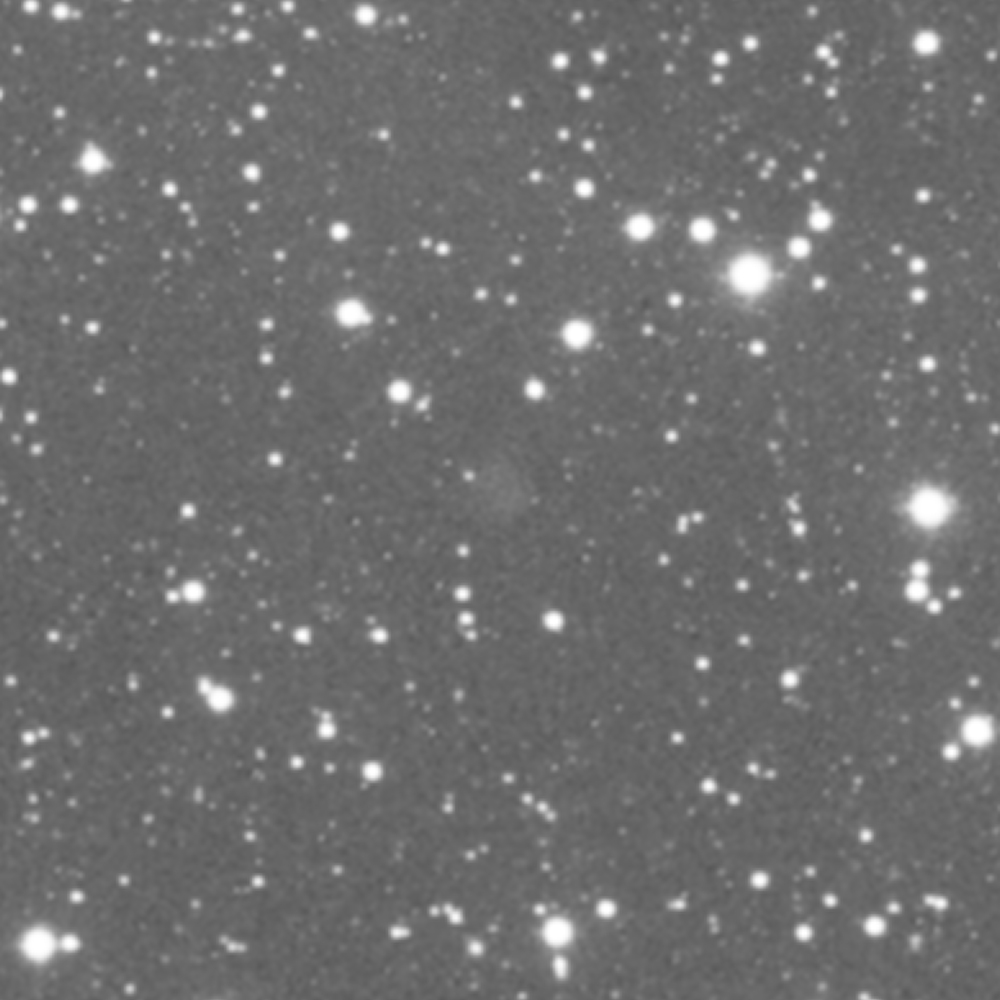}
  \caption{Discovery image of Donatiello I dwarf galaxy, obtained with a 127mm ED doublet refractor f/9 reduced at about f/6 with a Green filter and a 2MP cooled CCD camera from the Pollino National Park in Southern Italy. The total exposure time is about 6000 seconds obtained by combining (sum and median) images captured on 5,6,7 November 2010 and 5 October 2013 with the same equipment. The  total field of view is this cropped version centered in Do~I is 20$\times$ 20 $\arcmin$. North is up, East is left. \label{fig:error}} 
\end{figure}

Although the M31 stellar halo photometry from the PAndAS is significantly deeper than the SDSS, its larger distance makes it prohibitive to resolve the member stars of their companions fainter than the red clump level. This means that the satellite population (and its stellar streams) can only be traced by means of the less numerous red-giant branch (RGB) stars. The lack of enough RGB tracers in the CMDs of dwarf galaxies with absolute magnitude fainter than $M_{V}\sim -6$  makes it very hard to detect these lower-mass systems, yielding a cut-off in the luminosity function of satellites of M31 (see Brasseur {\it et al.} 2011; their Fig. 1).

An alternative searching method is the detection of unresolved (or partially resolved) stellar systems as diffuse light structures in low-resolution, deep images taken with small telescopes (Mart\ \i nez-Delgado et al. 2016; Besla et al. 2016) in the outskirts of the Local Group spirals. Because of the independence of the surface brightness on the distance in the nearby universe, this  alternative approach also  allows the detection of systems beyond the Local Group, up to distances of 10-15 Mpc (e.g. Danieli, van Dokkum \& Conroy 2018). Thus,  there is still room for discovery of low surface brightness dwarf galaxies  which may lurk in the regions where the detection efficiency drops to very small values (e.g., Koposov et al 2008; see also Whiting et al. 2007 for a detailed discussion of the completeness of the nearby dwarf galaxy population based on diffuse light detection studies).

The diversity of dwarf galaxy properties in the local universe suggests that they have multiple formation paths (Lisker 2009). Some may have formed at early times in the Universe (e.g. Nagashima et al. 2005; Valcke et al. 2008; Bovill \& Ricotti 2009). Others are believed to have been transformed from spiral or irregular galaxies by environmental processes, such as tidal interactions, harrassment, ram pressure stripping, resonant stripping etc.\ (Moore et al. 1996, Mayer et al. 2001; Mastropietro et al. 2005; Lisker et al. 2006, 2007; D\'\ Onghia et al. 2009). An important way to disentangle environmental from intrinsic processes is to study nearby dwarf galaxies that are isolated from both nearby large neighbors and groups of other dwarfs.

\begin{figure*}
	\includegraphics[width=0.51\textwidth]{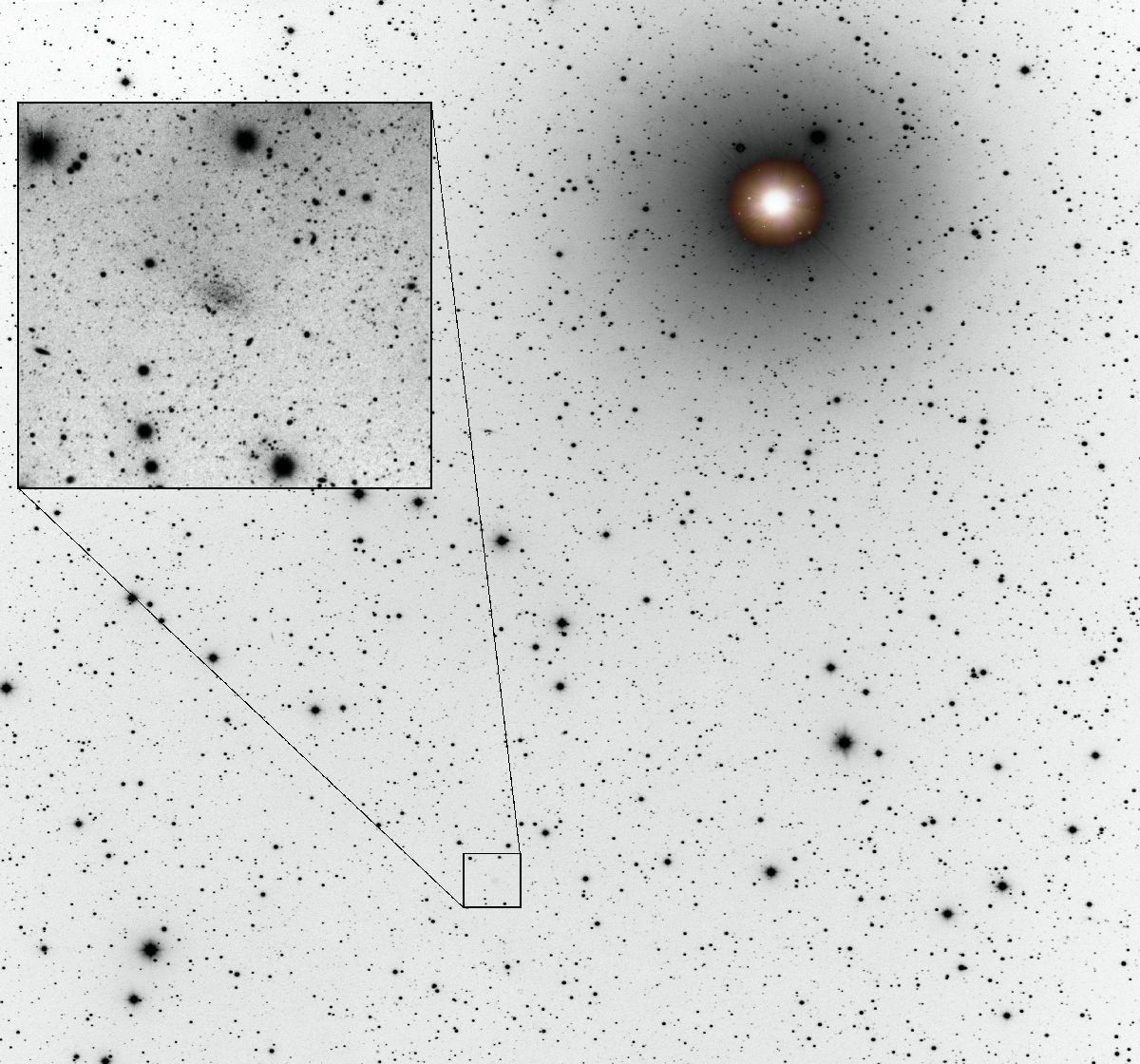} 
	\includegraphics[width=0.475\textwidth]{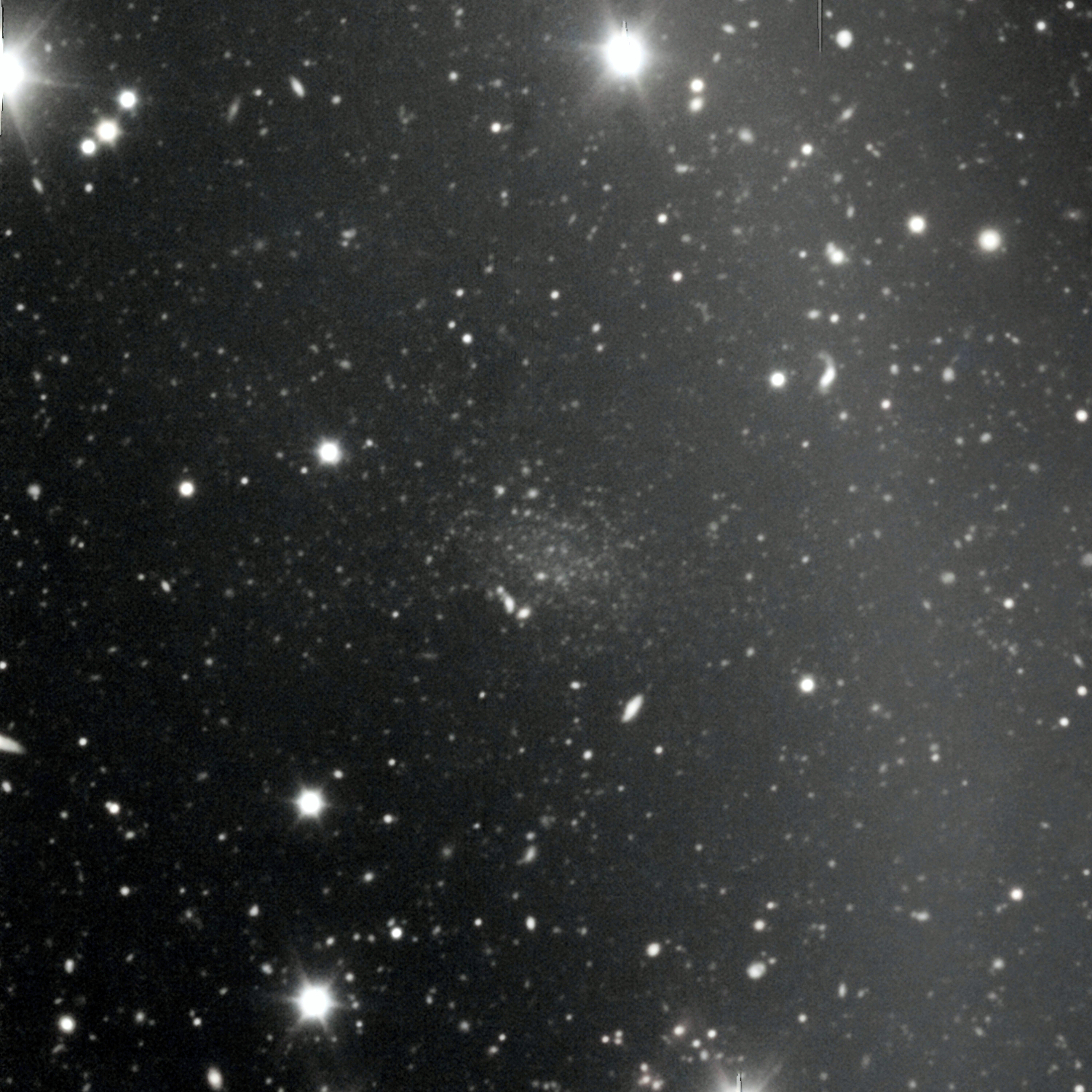}

    \caption{{\it Left panel}: Image of the dwarf galaxy Do~I taken with a Takahashi FSQ106  refractor at f/3.6 and a commercial CCD camera QHY90A on 27 November 2016 at the Avalon Merlino Remote Observatory. The total exposure times were  24300 sec in the Luminance filter.The field of view of this cropped image is 123$\times$ 123 $\arcmin$ with a pixel scale of 2.89 arcsec~pixel$^{-1}$ (Credits: Matteo Collina \& Giuseppe Donatiello). The inner zoomed panel shows the TNG $g$-band image of the galaxy (see Sec. 2.1). {\it Right panel}: $r$-band image of Do ~I obtained from GTC observations (see Sec. 2.2). The total field of view is 4.30$\arcmin$ $\times$ 4.30$\arcmin$. North is up, East is left}
    \label{fig-cmd}
\end{figure*}

The existence of  isolated dwarf spheroidal (dSph) galaxies is predicted by some
direct hydrodynamical simulations, but the true nature of these
systems remains an open question. The transformation of
star-forming, gas-rich dwarf irregular galaxies (dIrr) to gas-poor,
passive dSphs is most commonly attributed to the effects of tidal and
ram-pressure stripping when a previously isolated dwarf galaxy becomes
a satellite of a larger system (e.g. Grebel, Gallagher \& Harbeck 2003; Mayer 2010). Dwarfs that form in close
proximity to massive halos may furthermore be stripped by local reionization
processes and internal heating (e.g. Ocvirk \& Aubert, 2011; Nichols 
\& Bland-Hawthorn 2011) However, other mechanisms exist that can quench star 
formation and result in dwarf spheroidal galaxies independent of environment. 
The very shallow potential wells and low virial temperatures of galaxies in halos with
dynamical masses below $\sim 10^9 {\rm M}_{\odot}$ can lead to an almost complete
depletion of the interstellar medium due to a combination of supernova
feedback and UV background radiation (e.g., Sawala, Scannapieco \& White 2011;
Simpson et al. 2013).  As shown by Revaz et al. (2009), such systems can have
star formation histories and stellar populations very similar to those
of observed dwarf spheroidals near the MW. A further mechanism that
could result in presently isolated dwarf spheroidal galaxies was
recently demonstrated by Benitez-Llambay et al. (2013), who showed that
gas can be stripped from a low-mass halo via ram-pressure as it is
passing through the cosmic web at high redshift.

However, such isolated dwarf spheroidals (dSph) are extremely rare in the Local Volume. Geha et al. (2012)  found that quenched galaxies with stellar masses of $\sim 10^9 {\rm M}_{\odot}$ and magnitudes
 of $M_r> -18$ (properties that include dEs and dSphs) do not exist beyond 1.5 Mpc from the nearest host ($\sim 4$ virial radii). 
 Indeed, only a few distinct cases of isolated dSph galaxies have 
been discovered so far: KKR~25 \cite{2001A&A...379..407K,2012MNRAS.425..709M} at 1.9 Mpc, KKs~3 \cite{2015MNRAS.447L..85K} at 2.1~Mpc and Apples~1 \cite{2005AJ....129..148P} at 8.3 Mpc have no known galaxies closer than 1 Mpc to them. Karachentseva et al. (2010) presented a list of 85 very isolated dwarf galaxy candidates within 40 Mpc, ten of which were classified as dwarf spheroidals. However, these authors pointed out the difficulties in finding isolated dSphs that cannot be detected in the 21 cm line surveys and remain undiscovered due to their very low surface brightness. Makarov et al. (2015) carried out spectroscopic observations of three galaxies from the list of isolated dSph galaxies and concluded that they are not isolated systems. The authors noted that only two (KK~258 and UGC~1703) of the seven studied so far candidates are isolated dSph galaxies. The transition-type dwarf galaxy KK~258 \cite{2014MNRAS.443.1281K} resides at 0.8~Mpc from Sdm galaxy NGC~55. According to the Local Volume database \cite{2013AJ....145..101K} UGC~1703 has no companions closer than 0.8~Mpc.
The number of known very isolated dSph systems does not exceed 5 of more than 1000 of the galaxy population in the Local Volume. The study of such isolated dwarfs populated exclusively by old stars is of great interest to investigate the influence of their environments vs. internal processes on their star formation histories, and to test models that assume a different origin for these galaxies.

 In this paper, we report the discovery with a small telescope of a  dwarf spheroidal galaxy projected on a halo field of M31, which was previously missed by surveys based on resolved star counts and on the H{\sc i} 21-cm radio line. It was found as part of a new project to search for faint dwarf  satellites in the Local Group with amateur telescopes.

\section{OBSERVATIONS AND DATA REDUCTION}

Donatiello ~I (Do~I) was first found on 23 September 2016 in a visual inspection of a
deep amateur image of the Andromeda galaxy region  obtained from the composition of a set of images taken during several years with a  refractor ED127mm (with an aperture of 12.7-cm at focal ratio $f/9$) by the amateur astronomer Giuseppe Donatiello (see Fig.1). The main purpose of this wide-field mosaic of the area between M31 and M31 was to detect the stellar streams and M31 satellites previously reported in the PAndAs as diffuse light structures using small amateur telescopes. The detection was confirmed by a visual inspection of the SDSS DR9 images and follow-up observations using the professional facilities described in Sec. 2.1 and Sec. 2.2. The position of the center of this new dwarf galaxy is given in Table~1.

\subsection{TNG imaging observations}

We used deep images of the field around the dwarf galaxy obtained with
the instrument Device Optimized for the Low Resolution (DOLoRes) of
the 3.58-m Italian Telescopio Nazionale Galileo (TNG)(Roque de Los Muchachos Observatory, La Palma, Spain)
taken on November 27 2016. These observations include 20x180 sec
exposures in the $g'$-band and 15$\times$180 sec exposures in the $r$'-band, with a seeing of 0.8"--1" and $\sim$~1", respectively. The total field-of-view was 8.6'$\times$8.6' with a pixel scale of 0.252".

The preprocessing of the data was performed in a standard way using
IRAF tasks, i.e. dividing the trimmed and bias-subtracted images by a
master flat field produced from multiple twilight sky-flat
exposures. For each filter, a final image for photometric processing
was obtained from a median stack of the preprocessed images. 

The instrumental magnitudes 
were derived using \textsc{daophot ii} and  \textsc{allstar} (Stetson 1987) packages. Sources
were detected in the $g$-band by requiring a 3-$\sigma$ excess above
the local background, with the list of $g$-band detections  being 
subsequently used for the \textsc{allstar} run on the $i^\prime$-band image. 
To obtain clean CMDs and minimize the contamination by faint background galaxies, 
 sources with the ratio estimator of the pixel-to-pixel scatter $CHI <$ 2.0 and sharpness parameters
$|S| < 1.0$ (see Stetson 1987) were kept for 
further analysis.  The instrumental magnitudes were calibrated and transformed
to the SDSS system using stars from the SDSS
DR13 catalog. For this calibration we used the pipeline developed and used in Javanmardi et al. (2016) and also used in Helkel et al. (2017). First, SExtractor (Bertin \& Arnouts 1996) was run on the image and all the objects in the image were detected and their flux was measured. Then the coordinates of the detected objects were fed into the SDSS Data Release 13 (Albareti  et al. 2016) CrossID and only the stars with $r\geq15$ mag, $0.08<(r-i)<0.5$, and $0.2<(g-r)<1.4$ were selected. Then a series of consecutive linear fitting ($r_{sdss}$ vs. $r_{image}$ and $g_{sdss}$ vs. $g_{image}$) with outlier rejection was performed. In each fitting round the stars with $\Delta \equiv ($mag$_{SDSS}$-mag$_{calibrated})>2\sigma$ were rejected, where $\sigma$ is the standard deviation of $\Delta$. The procedure was repeated until no $2\sigma$ outliers remained. 
 The final standard deviations in $\Delta$ are 0.043 and 0.015 mag for $r$ and $g$ bands, respectively. They were added in quadrature to the uncertainties of the subsequent measurements.

\subsection{GTC imaging observations}


We took deep images of the Do~I dwarf galaxy on January 18 and January 24, 2017 with the Optical System for Imaging and low-Intermediate-Resolution Integrated Spectroscopy (OSIRIS)\footnote{\url{http://www.gtc.iac.es/instruments/osiris/}} installed on the 10.4-m Gran Telescopio Canarias (GTC) (Roque de Los Muchachos Observatory, La Palma, Spain) using an unvignetted field of view (FOV) of 7.8'$\times$7.8' and a scale of 0.254" pixel$^{-1}$. The total exposure time was 22 $\times$ 150 sec =3300 sec for each of the three $g'$,$r'$, and $i'$ photometric bands. An 11-point dithering pattern (2$\arcsec$ offsets) was used to correct for chip defects and improve image sampling. The seeing was around 1.1$\arcsec$ and 0.9$\arcsec$ in $g'$ on January 18, and around 1.1$\arcsec$ and 0.9 $\arcsec$  in the $g'$ and $r'$-band on January 24, respectively. The resulting $r$-band  image is shown in Figure~2({\it right panel}).

\begin{figure}
\begin{center}
\includegraphics[width=0.50\textwidth]{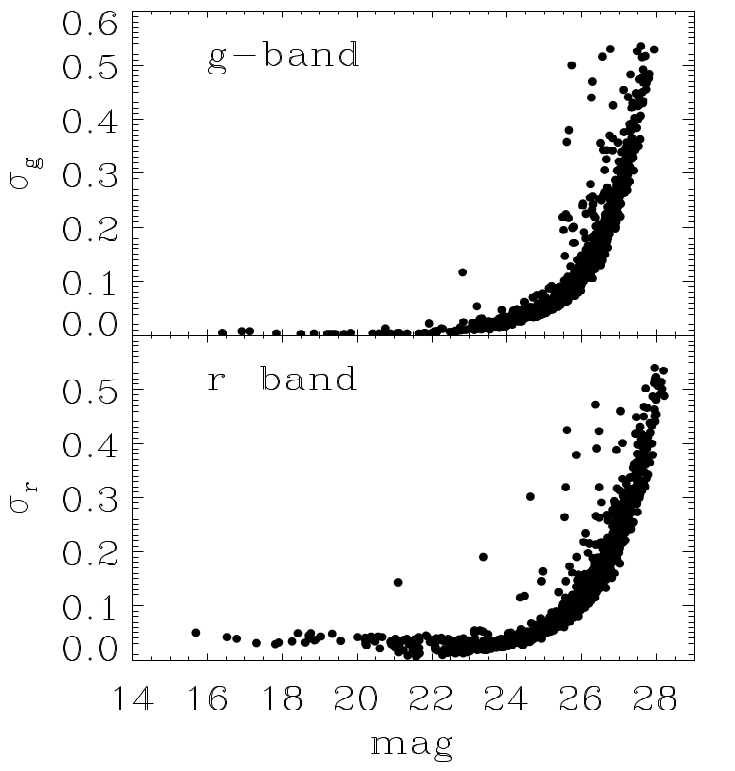} 
\includegraphics[width=0.475\textwidth]{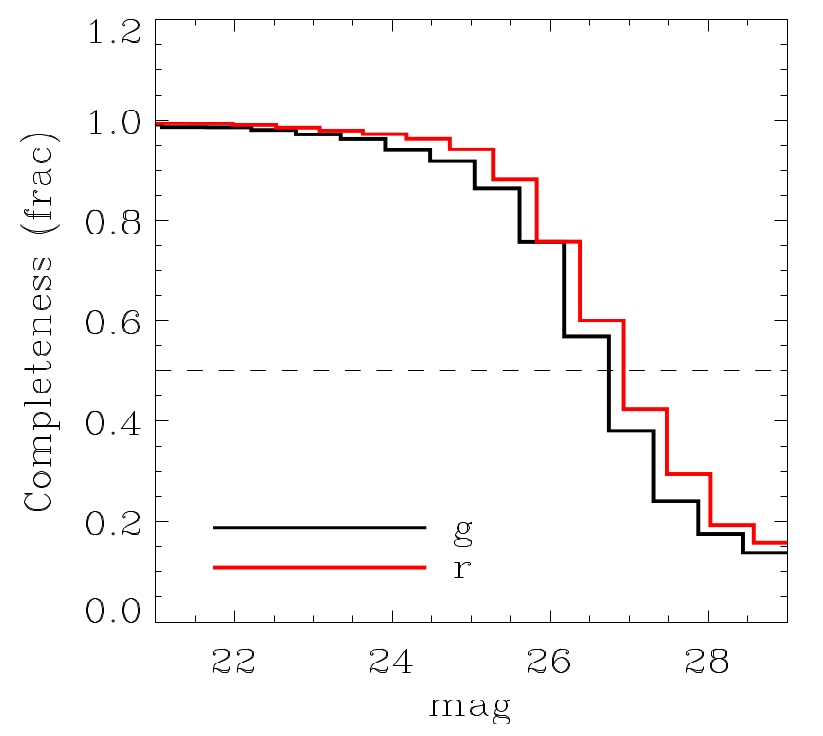}
  \end{center}
  \caption{{\it (Upper panel)}: Photometric  error  parameter for the resolved stars found  DAOPHOT II/ALLSTAR in our GTC images as a function of magnitude in the $g$ (top) and $r$-bands (bottom), including only sources with
sharpness parameter between -0.5 and 0.5. {\it (Lower panel)}: Fraction  of  artificial  stars  recovered as function of
input magnitude in g (black line) and r-bands (red line).  The
horizontal dashed line marks the magnitude corresponding to a completeness of 50\% (see Sec. 2.2) .\label{fig:error}} 
\end{figure}

We pre-processed the scientific images using IRAF tasks in the standard way (i.e. performing over-scan correction, bias subtraction and flat field division). In particular, we used the OSIRIS Offline Pipeline Software (Ederoclite \& Cepa 2010), a python script that calls IRAF routines in a user-friendly graphical user interface. For each filter, we produced a final image for photometric processing by doing a median stack of the preprocessed images.

 Photometry was obtained with our own pipeline based on the PSF-fitting algorithm of \textsc{DAOPHOT II/ALLSTAR} (Stetson 1987). The final catalog only includes stellar-shaped objects with |sharpness| < 0.5. For the calibration of the images in the $g$ and $r'$ bands we use the same semi-automatic method introduced in Javanmardi {\it et al.} (2016) and briefly described in Sec. 2.1.  
 The final standard deviations in $\Delta$ are 0.031 and 0.045 mag for the $r$ and $g$ bands, respectively, and were added in quadrature to the uncertainties of the subsequent measurements.

In order to estimate the completeness of our photometric catalogs, we inserted synthetic stars in the images using \textsc{DAOPHOT\,II}. The number of artificial stars was 10\% of the observed sources for each of the frames. These artificial stars were randomly distributed throughout the chip. The experiment was designed to only include stars with magnitudes in the range $17 < g,r < 28$. PSF-fitting photometry was derived for these altered images by applying the same PSF model used for the observed stars. We then measured the fraction of synthetic stars recovered by our procedure and the mean variation  of the completeness as a function of the magnitude was derived. Our photometry recovers nearly all the synthetic stars up to $g', r' \sim$ 24 and drops below 80\% at $g',r' \sim$ 25.7. We set the limiting magnitude of our photometry, with a completeness of 50\%, at mag = 26.7 and 27.0 for the $g'$ and $r'$ bands, respectively (see Figure~3, lower panel).



\begin{table}
\begin{center}
\caption{\label{properties}Properties of the Donatiello~I dwarf galaxy}
\begin{tabular}{l|c}
\hline
\hline

$\alpha$ (J2000) & $01^\mathrm{h}11^\mathrm{m}40.37^\mathrm{s}$\\
$\delta$ (J2000) & $+34^o36'03.2''$\\
$\ell$ & $127.65^\circ$\\
$b$ & $-28.08^\circ$\\
$(m-M)_0$ & $27.6\pm0.2$\\
Heliocentric Distance & $3.3\pm 0.1$ Mpc\\
$M_{V}$ & $-8.3^{+0.3}_{-0.3}$\\
$L_{V}$ & $10^{5.3^{+0.2}_{-0.3}}$\(L_\odot\)\\
$\mu_{V}$ (mag arcsec$^{-2}$)& $26.5 \pm 1.0$\\
Ellipticity & $0.69 \pm 0.05 $\\ 
$r_{h}$ & $0\farcm48^\pm 0.15$\\
& 442$\pm$157 pc\\
\hline
\hline

\end{tabular}
\end{center}
\end{table}

\section{RESULTS}
\label{sec:maths} 

\subsection{Distance}

\begin{figure}
\begin{center}
\includegraphics[width=0.50\textwidth]{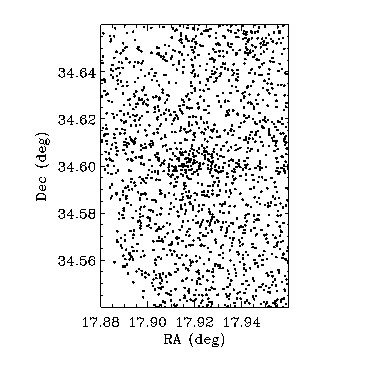} \includegraphics[width=0.475\textwidth]{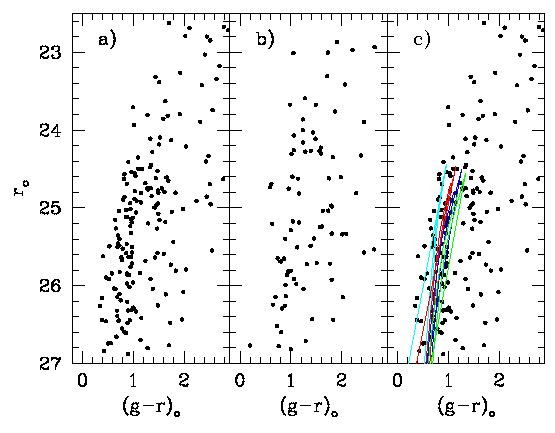}
  \end{center}
  \caption{{\it (Upper panel)}: Star count map of Do~I showing  the stars detected in both $g$ and $r$-bands in our GTC observations. {\it (Lower panel)}: De-reddened $(g-r)_0$ vs. $r_0$ color-magnitude diagram of Do~I (panel a) and its control field (panel b). Panel c) shows a comparison of the Do~I CMD with the PARSEC isochrones (solid lines) with an age of 13 Gyr shifted assuming a distance modulus of $(m-M)$=27.6 mag. The metallicities of the isochones  are 
  $Z$=0.0001 (cyan), 0.0003 (red), 0.0005 (blue) and 0.0007 (green) respectively.\label{fig:galfit}} 
\end{figure}

Figure~4 (lower panel) shows the color--magnitude diagram (CMD) of Do~I for  a selected region of 1.8 $\arcmin \times$ 1.8 $\arcmin$  from the GTC observations (panel a). The inner region of Do~I (within a radius of 0.5 $\arcmin$ centered in the galaxy) is seriously affected by the crowding and was excluded from this diagram.\footnote{See  Mart\'\i nez-Delgado \& Aparicio (1997) (e.g. their Fig.1) for a comprehensive discussion of the crowding effects in the observational errors and the  CMD morphology gradient for the case of an old-population dwarf galaxy.}
The main feature of the diagram is the red giant branch (RGB) locus, suggesting that the stellar population of this dwarf galaxy is dominated by old, metal-poor stars similar to those observed in dSph galaxies in the Local Group.  Although we selected a narrow field of view centered on the dwarf, there is still contamination by brighter RGB stars from the halo of M31 (and some Galactic foreground stars too) that overlap with this feature in the CMD, as it is also showed in the CMD of the control field situated $\sim$3.5$\arcmin$ North from the center of the galaxy (Fig. 4, panel b). However, the stellar density maps for different magnitude cut-offs  show no detection of the dwarf for $r$-band magnitude $r >$ 24, suggesting the bulk of the red stars of this galaxy are fainter than that magnitude. This suggests that this partially resolved system is beyond the Local Group and is not a satellite of the Andromeda galaxy. This also explains its absence in the list of over-density candidates detected with automatic detection algorithms in the PAndAs survey (Martin et al. 2013). A visual inspection of the images from this survey  shows Do~I was only visible as a partially resolved object.

The low number of detected upper RGB stars and the M31 halo star contamination makes difficult to estimate the position of the tip of the RGB (TRGB) in the diagram (e.g. by means a Sobel filter; Madore \& Freeman 1995). For this reason, we attempted to estimate the distance of Do~I by the means of CMD fitting following the technique presented in Longeard et al. (in preparation), which models the distribution of stars in the CMD as the sum of a single stellar population and a constant field contamination, estimated from the edge of the field of view. Unfortunately, both the dwarf galaxy's CMD and the field contamination model proved to be too noisy to yield sensible results. In particular, we were not able to determine any reliable distance from this analysis, despite testing distance moduli in the 24.0-28.0 range. An attempt at fitting only stars fainter than $r_0$ = 23.5 was made in order to discard potential bright stars from the M31 halo that could affect the fit, but the results were still inconclusive. Nevertheless, the distribution of stars in the CMD of Do I leaves no ambiguity that Do I stars are present in the field, as also shown in the stellar density map built from our photometry in Figure ~4 (upper panel).  Thus, an accurate distance to Do~I cannot be obtained from the current data with this approach.

\begin{table}
 \begin{center}
  \caption{The effective $V$-band surface brightness, $\mu_e$, and total $V$-band magnitude measured by GALFIT, the S\'ersic index, $n$, the effective radius, $R_e$, and the axis ratio, $b/a$. The physical parameters in the lower part of the Table (i.e. effective radius in pc, absolute $V$-band magnitude, $M_V$, and the luminosity) are calculated by assuming the distance modulus given in Table 1. The $V$-band quantities were obtained by transforming $g$ and $r$ band magnitudes to $V$ magnitudes using  Fukugita et al. (1996).\label{t:galfit_results}}
  \begin{tabular}{ccc}
  \hline
   &GALFIT & SRP\\ 
   \hline
   \hline
   $\mu_{e,V}$ (mag.arcsec$^{-2}$)& $26.9 \pm 0.07$ & $26.5\pm 1.0$\\
   $V$ (mag) & $18.25 \pm 0.07$  & -\\
   $n$ & $0.73 \pm 0.02$ & -\\
   $R_e$ (arcsec) & $20.7\pm 0.3$ & $28.8 \pm 9$ \\
   $b/a$ & $0.68 \pm 0.01$ & 0.69 $\pm$ 0.05 \\
   \hline
   \hline
   $R_e$ (pc) & $332 \pm 44$ & $442 \pm 157$ \\
   $M_V$ (mag) & $-9.3\pm0.3$ & $-8.3^{+0.3}_{-0.3}$\\
   $\log_{10}\left(\frac{L_V}{L_{\odot}}\right)$ & $ 5.7\pm0.1$ & $5.3^{+0.2}_{-0.3}$\\
    \hline
  \end{tabular}

 \end{center}
\end{table}

 \begin{figure}
\begin{center}
 \includegraphics[scale=0.8]{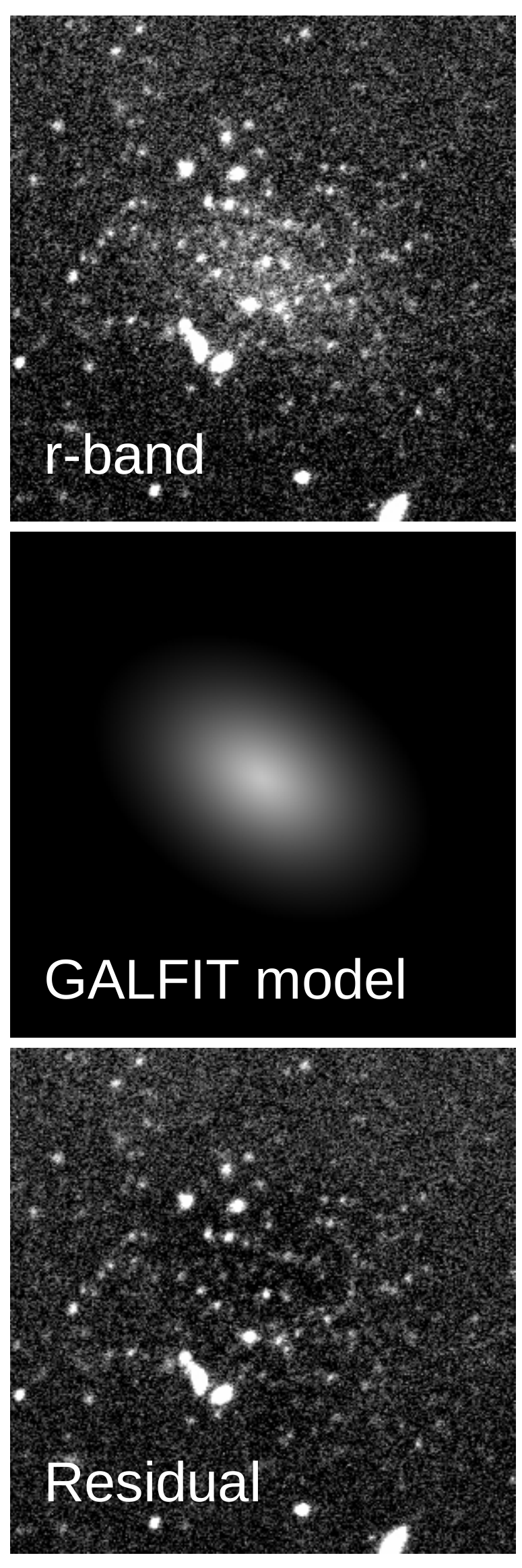}
  \end{center}
  \caption{GALFIT modelling of Do~I from the TNG $r$-band image of Do~I. From top to bottom: the original image centered on the dwarf, the GALFIT fitted model, and the residual image obtained by subtracting the model from the original image. \label{fig:galfit}}
\end{figure}

\begin{figure*}
\begin{center}
 \includegraphics[scale=0.7]{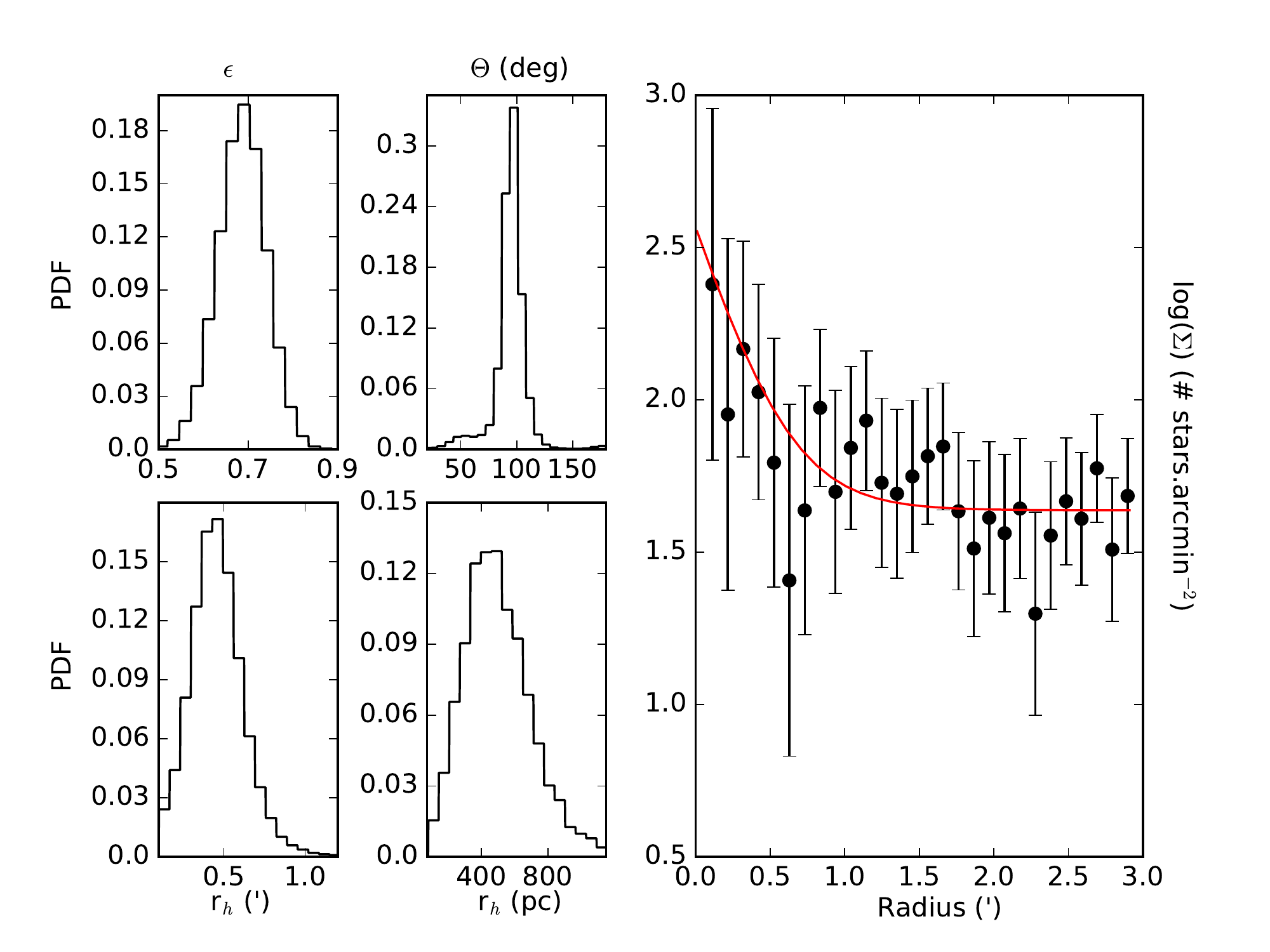}
  \end{center}
  \caption{{\it Left} : Probability distribution functions for some of the structural parameters of Do~I from the Resolved Stellar Population (RSP) method: the half light radius r$_{h}$ in angular and physical size, the position angle $\theta$ and the ellipticity. {\it Right} : Radial density profile within 3 $r_{h}$ of Do~I centroid. The number of stars per unit area is represented as a function of the galactocentric radius in arcminutes. The solid curve represents the exponential radial profile drawn from the favoured structural model.  \label{fig:galfit}}
\end{figure*}

Since the lack of a well-defined tip of the RGB  prevents us from estimating an accurate  distance, we explore the possibility that Do~I is associated with NGC 404, which is at a projected distance of only 73$\arcmin$ (see Sec. 4). For this purpose, we obtain a distance modulus for Do~I from an eye-ball isochrone fitting of the RGB locus position 
in its de-reddened CMD, assuming this galaxy is mainly composed by old stars. Galactic extinction values for Do~I of $A_{g}$=0.206 and $A_{r}$=0.143 (Schlafly \& Finkbeiner 2011) were used. 
Figure 4 (panel c) shows the PARSEC isochrones (Bressan et al. 2012) in the metallicity range 0.0001 dex ([Fe/H]=-2.2) to 0.0007 dex ([Fe/H]=$-1.3$)  and an age of 13 Gyr shifted for a distance modulus of $(m-M)=27.6 \pm 0.2$ (dist=3.3 Mpc). 
This value is consistent (within the uncertainty) with the distance moduli of NGC 404 reported in recent studies ($(m-M)=27.67 \pm 0.15$;  Tikhonov, Galazutdinova \& Aparicio 2003; $(m-M)=27.46 \pm 0.02$: Lee \& Jang 2016). Our isochrone comparison also suggests that a distance larger than $\sim$ 4 Mpc (i.e. $(m-M)>$ 27.8) is very unlikely for Do~I. 

Alternatively, there are several brighter red giant star candidates (in the range $23.7<r<24.5$)\footnote{Because of the low-quality of our CMD, we cannot confirm with the current data the presence of brighter, intermediate-age stars above the RGB locus (e.g. see Martinez-Delgado \& Aparicio 1997), which could also bias the distance estimate of Do~I.}  which could suggest a closer distance for Do~I. If the actual TRGB lies at the position of the brightest star in this group at $r_0 \sim 23.7$ (see Figure 4, lower panel), our isochrone comparison would lead to a lower limit for the distance of Do~I of 2.2 Mpc.  However, since these stars are probably foreground M31 halo stars, we can only assert that Do~I is located within a distance range of 2.5 to 3.5 Mpc, and thus a non-member of the Local Group of galaxies.

\subsection{Structural properties and surface brightness profile}

Because of the partially resolved appearance of Do~I  in our images (see Fig. 5, upper panel), we  have used two different approaches to measure its structural properties: i) a GALFIT model fit of the main body of the dwarf as an unresolved, diffuse light system as traced by the TNG data (see Sec. 2.1); ii) the fit of the main body as traced by the resolved stellar population from the GTC stellar photometry.

For the first approach, we fit a S\'ersic model using the GALFIT software \cite{2010AJ....139.2097P}. For this purpose, we first cut a sub-image of 400 pixels $\times$ 400 pixel of the image centred on the dwarf galaxy, mask the neighbouring foreground stars and background galaxies, and then apply GALFIT. The results are shown in Figure~5 and Table~2. 
Fig. 5 shows the 400 pixels $\times$ 400 pixels r-band image centred on the dwarf galaxy, the Sersic model obtained by GALFIT, and the residual image, which was obtained by subtracting the model from the original image. We note that some resolved stars can be seen in the residual image, which means that they were not taken into account in the GALFIT modelling.
Although these objects are clearly brighter than the Do~I  stars detected in the CMD (see Sec. 3.1) and thus are M31 halo star candidates, we cannot reject the possibility that
 the brightness of the dwarf galaxy was slightly underestimated by GALFIT and the surface brightness profile is also not exact.

Alternatively, to derive the structural parameters of the dwarf galaxy, we use the $g$ and $r$ bands  
following the technique of Martin et al. (2016) and Longeard et al. (in prep), which  accounts for gaps in the spatial coverage of the data. Stars with photometric uncertainties greater than 0.2 are discarded in both bands. We assume an exponential profile for Do~I and a flat background contamination across the field. The analysis yields posterior probability distribution functions (PDFs) for six parameters : x0 and y0, offsets from a chosen centroid, which are easily converted to the system's centroid; $r_{h}$, the half-light radius (along the major axis) of the dwarf galaxy; $\epsilon$, 
its ellipticity\footnote{The ellipticity $\epsilon$ is defined here as $r_{h}$ = $b$  ($1-\epsilon$), with $b$ the minor axis of the ellipse.}; 
$\theta$, the position angle of the major axis east to north; and $N_*$, the number of stars within the chosen CMD selection box. The PDFs for $r_{h}$, $\epsilon$ , and $\theta$ are shown in Figure 6 and all the structural parameters are summarized in Table 2. 
Finally, we work under the assumption that the Do~I half-light radius should not be smaller than 0.1 arc minutes on the sky.

The radial density profile of the dwarf galaxy is displayed in Figure 6 and compared to the star counts in elliptical annuli. The theoretical exponential profile drawn from the favoured structural parameters is also shown in Figure 6, in red, and shows good agreement with the binned data.

Our method to estimate the luminosity follows the analysis described in Martin {\it et al.} (2016) and can be interpreted as the statistical determination of the total luminosity of any system with the same structural properties as Do~I.
At each iteration, a number of stars $N_{*,j}$ is randomly drawn following the $N_{*}$ posterior distribution function obtained through our structural fitting procedure. The distance modulus is also randomly drawn from a Gaussian distribution of mean $(m - M)$ = 27.6 and standard deviation $\delta_{(m-M)}$ = 0.2. We also choose a stellar population of [Fe/H] = $-2.0$, 12 Gyr and solar alpha abundance, which corresponds to the typical population one should expect from a MW/M31 dSph with Do~I's total luminosity (Kirby {\it et al.} 2013). The choice of a slightly different age, metallicity or [$\alpha$/Fe] would not impact the inference of the luminosity of the system. A star is then simulated along this isochrone : its $g_0$ and $r_0$ magnitudes are checked : first, they should fall within the magnitude and color limits of the data  (i.e. the min and max $g_0 - r_0$ and $g_0$ in our CMD). For simulating the completeness in the CMD, two random numbers $a$ and $b$ between 0 and 1 are randomly drawn and respectively  compared with the completeness at the $g_0$ and $r_0$ of the simulated star. If the completeness is greater than $a$ and $b$, the star is flagged. A total number of $N_{*,j}$ are accumulated in this way. Finally, the fluxes of all simulated stars, flagged or not, are summed to give the total luminosity of the system. This procedure is repeated several times in order to obtain a PDF in luminosity and absolute magnitude.

Table 2 shows a comparison of the results for the structural parameters obtained with these two different methods. The upper part of Table~1 gives the effective surface brightness, $\mu_e$, total apparent magnitude, the Sersic index, $n$, the effective radius, $R_e$, and the axis ratio, $b/a$. The physical parameters in the lower part of Table~2 (i.e. effective radius in pc, absolute magnitude, $M_V$, and the luminosity) are calculated by assuming the  distance modulus ($(m-M)_0 =27.6 \pm 0.2$ mag) for the dwarf galaxy. The resulting absolute magnitudes, surface brightness and
effective radii are also plotted in Figure 7, in comparison with those for
known Milky Way and M31 dwarf galaxy companions. Interestingly, the structural
parameters obtained from the resolved stars and diffuse light component yield
similar results, suggesting that Do~I has an absolute magnitude, surface brightness, and size similar to some of the ``classical" dSph companions of the Milky Way (e.g. Draco or Ursa Minor). Both methods provide the same value for its ellipticity, confirming this galaxy is 
very elongated as previously noted in the stellar density map plotted in Fig. 4 (upper panel).


\begin{figure}
 \includegraphics[scale=0.25]{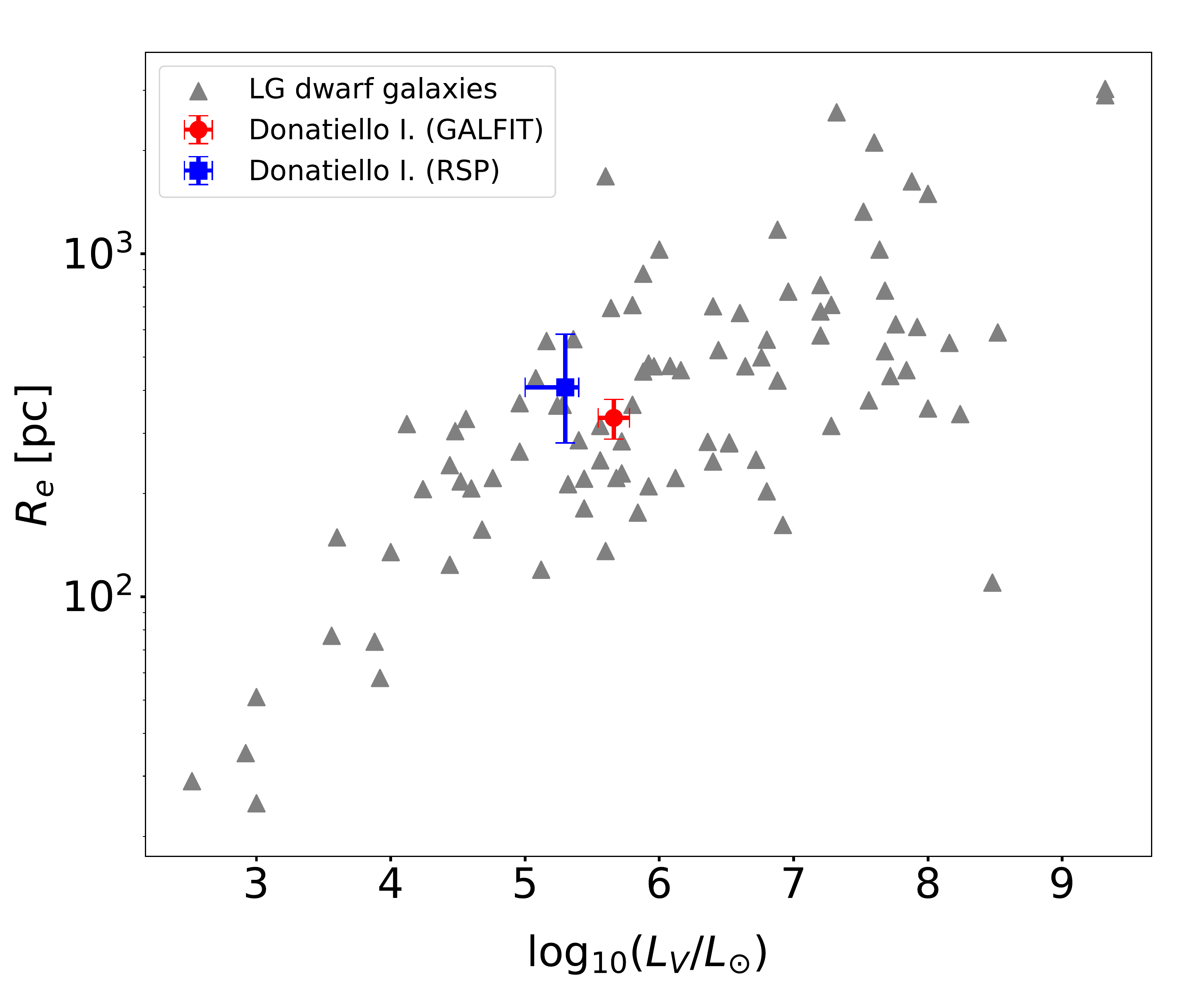}\\
 \includegraphics[scale=0.25]{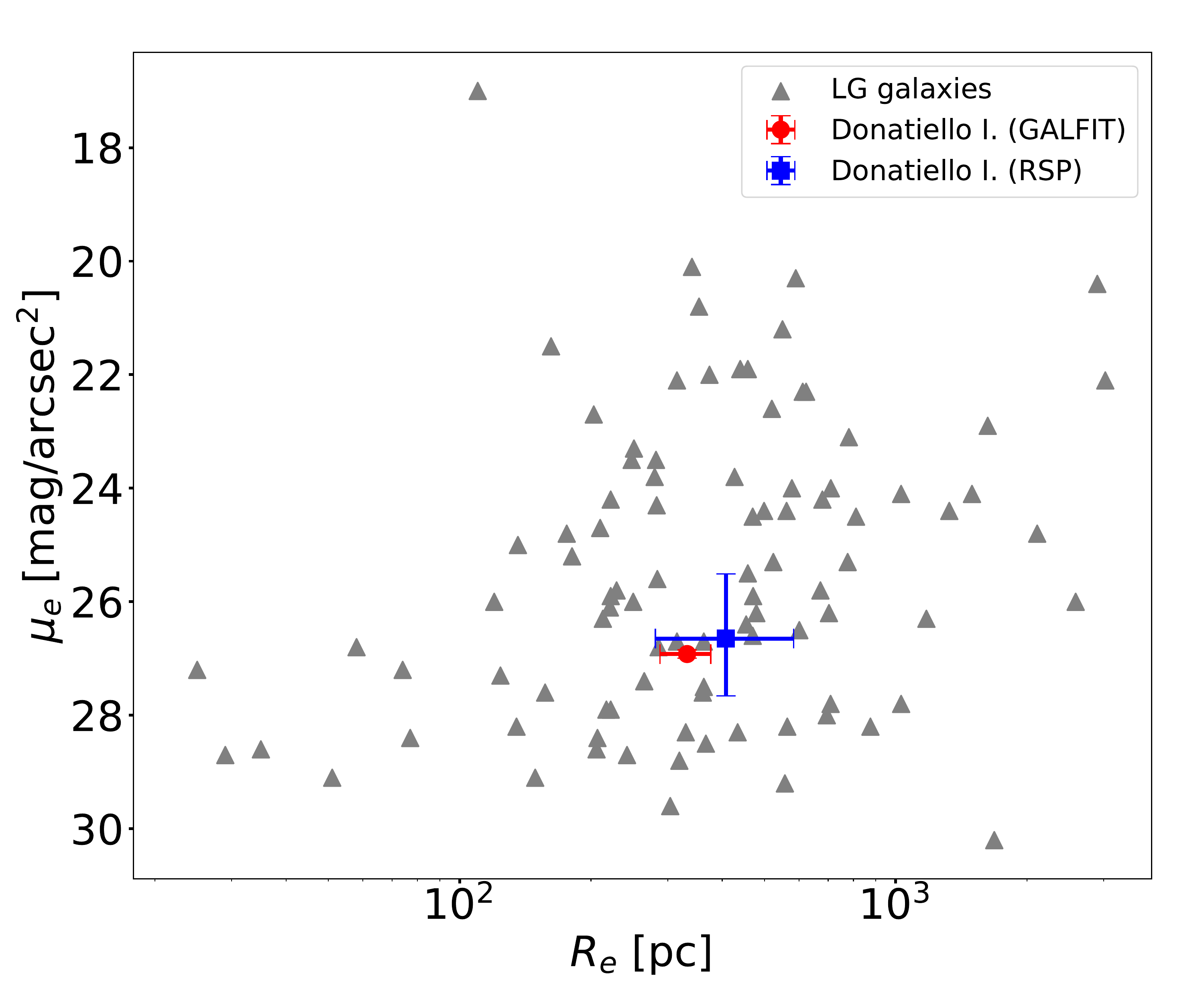}\\
 \includegraphics[scale=0.25]{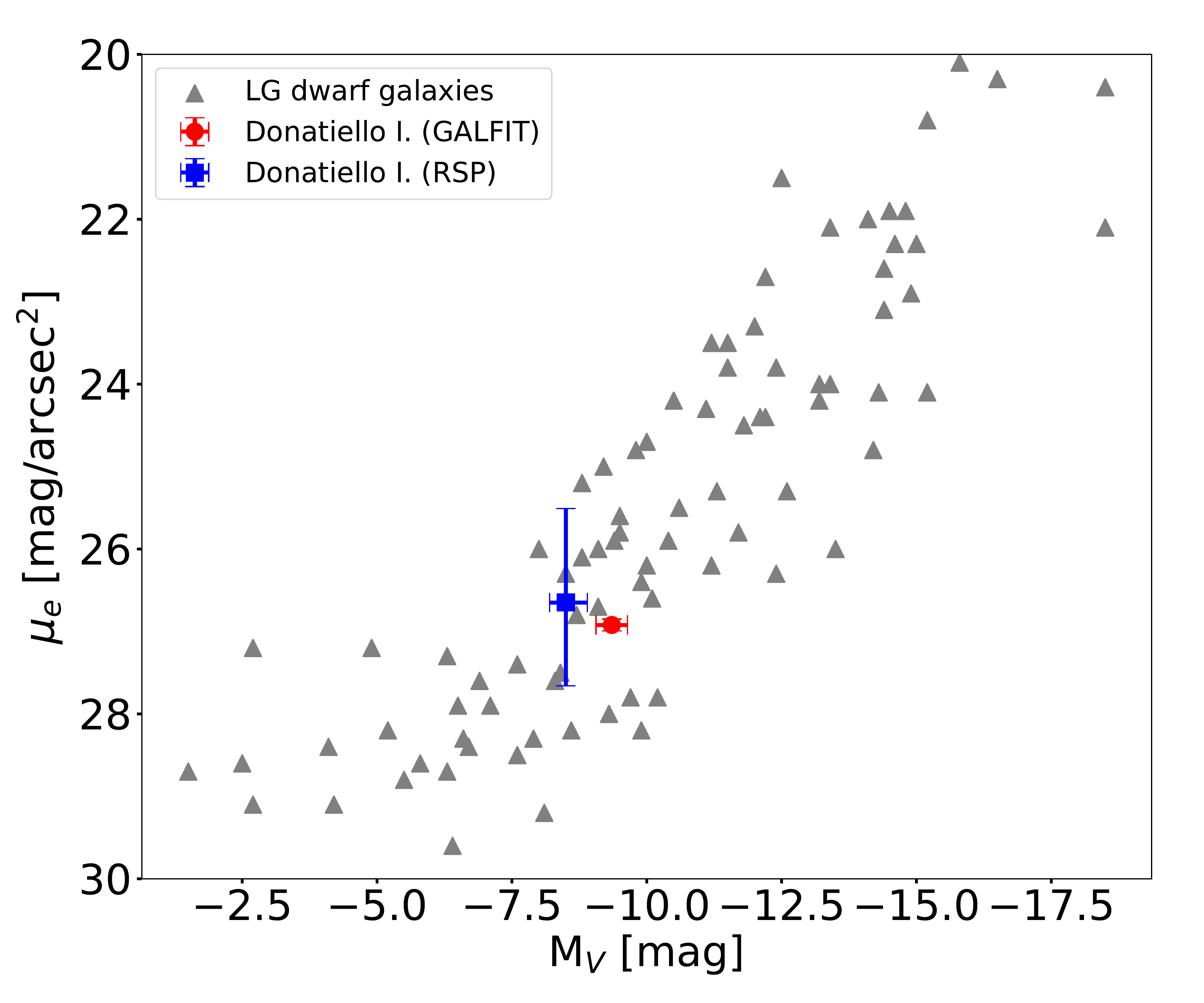}
  \caption{A comparison between the properties of Do~I and the known dwarf galaxies of the Local Group (McConnachie 2012). {\it Top}:  effective radius in pc vs.\ the logarithm of $V$ band luminosity in  $L_{\odot}$. {\it Middle}: $V$ band surface brightness in mag arcsec$^{-2}$ vs. effective radius in pc. {\it Bottom}: $V$ band surface brightness in mag arcsec$^{-2}$ vs. absolute $V$ band magnitude. The $V$ band quantities were obtained by transformations given in Fukugita et al. (1996), and the effective radius is the average of the $g$ and $r$ band measurements. \label{fig:compare_LG}}
\end{figure}


\subsection{H\,{\sc i} emission}

Do~I was not detected by the ALFALFA survey (Haynes et al., in preparation) but we have examined the ALFALFA grid at the relevant location to derive an upper limit to the H\,{\sc i} flux. Derivation of an upper limit to the HI mass for an individual object, for example as is described by Spekkens et al. (2014), requires assumptions about the source extent and spectral definition. Given that the $g$-band half-light radius is 24", we expect the H\,{\sc i} emission to be significantly
smaller than the effective ALFA beam (4 $\arcmin$) so that the H\,{\sc i} would be unresolved. For comparison with Spekkens et al.(2014), we smooth the ALFALFA data to 15 km~s$^{-1}$ in velocity and examine the ALFALFA cube over an ellipse following the standard ALFALFA minimum region of 7 $\arcmin \times 7 \arcmin$  (Haynes et al. 2018 in prep.) over ranges of relevant velocities. The complication is that the noise increases in the presence of Galactic H\,{\sc i} line emission; the standard ALFALFA drift scan strategy is not optimal for mapping the Galactic H\,{\sc i} emission although moderately strong, (nearly) unresolved features are identified (Giovanelli  et al 2013; Adams et al. 2013). Analysis of the ALFALFA grid at the position of Do~I gives an r.m.s. noise per beam solid angle of 2.0 mJy at 15 km~s${-1}$ resolution. For a signal of that width, the 5-$\sigma$ upper limit of the integrated line flux density at 15 km~s$^{-1}$ is 0.15 Jy~km~s$^{-1}$, translating to an upper limit on the  H\,{\sc i} mass of 3.5 $\times 10{^4} D^2$ $M_\sun$ where $D$ is the distance in Mpc. For a distance of 3.3 Mpc, this translates to 3.8 $\times$ 10$^5$ $M_\sun$.


It should be emphasized that this limit assumes that the HI emission in Do~I is unresolved both spatially and spectrally by the ALFALFA survey, both of which are reasonable assumptions.

\begin{figure*}
\begin{center}
 \includegraphics[scale=0.6]{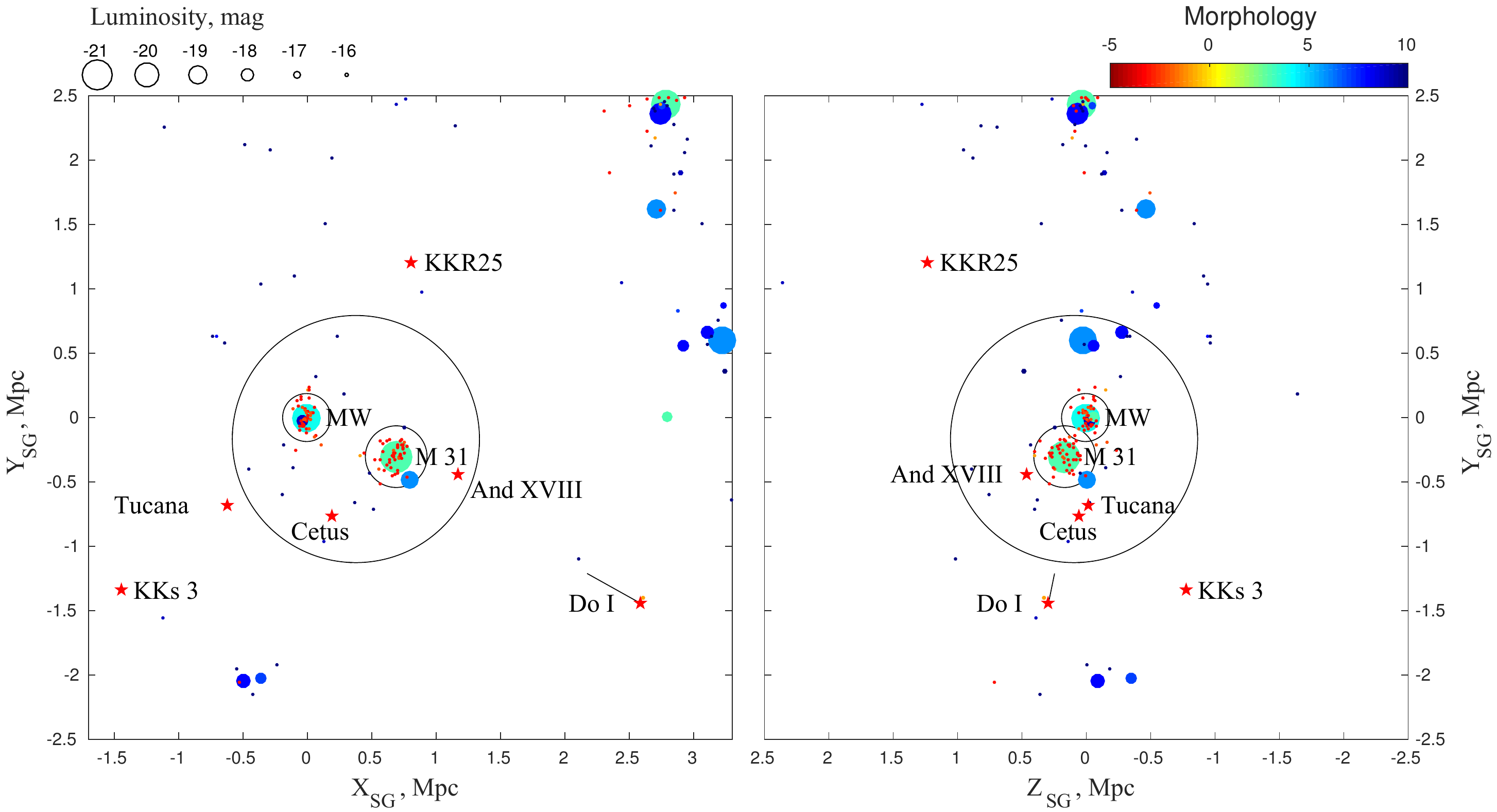}
  \end{center}
  \caption{ Super-galactic coordinate map of the distribution of galaxies within a 5~Mpc cube roughly centered on the Local Group. The size of filled circles is scaled to the luminosity of a galaxy. 
  The T-values morphological classification of each system is color-coded according to the color-bar scale. The big circle encompassing the Local Group corresponds to the zero-velocity surface \cite{2009MNRAS.393.1265K}, which separates the collapsing region of space around the Local Group from the cosmological expansion. Known highly isolated dSph systems are marked with a red star as well as Do~I. The short line from Do~I indicates the distance uncertainty from 2.5--3.5~Mpc. In this map, the position of NGC~404 approximately overlaps with those of Do~I at 3.5 Mpc.\label{fig:map}}
\end{figure*}

\section{DISCUSSION}

 We have reported the discovery of Do~I, a faint stellar system at a projected distance of one degree from the Mirach ($\beta$ And) star. Our data suggest a distance range of 2.5--3.5 Mpc for Do~I, and
 its structural parameters and absence of H{\rm I} are consistent with those  of a dwarf spheroidal galaxy, similar to the known companions of the Milky Way or M31.   

Do~I lies 72.4 arcmin on the sky from NGC~404, which corresponds to a projected separation of 65~kpc if we adopt the TRGB distance of NGC~404 equal to 3.1~Mpc (Lee \& Jang 2016). In that case, the absolute $V$-band magnitude of Do~I is $-8.2$~mag, which is about 9.6~mag fainter than NGC~404 itself, $M_V=-17.79$. Interestingly, NGC 404 has likely experienced a recent interaction/merger, leading to a kinematically decoupled core and an HI shell around the galaxy (del Rio et al. 2004; Bouchard  et al. 2010). Thilker et al. (2010) showed  ongoing star formation in an 
extended disk-like HI ``ring" around this lenticular galaxy visible in the GALEX images.  del Rio et al. (2004) argued that this gaseous ring around NGC 404 is the remnant of a merger with a dwarf irregular galaxy which took place some 900 Myr ago. The proximity of Do~I and its elongated shape lead one to  speculate that it might have been involved in that interaction or potentially have been affected by it. A more accurate distance and a radial velocity measurement are needed to shed more light on this interaction scenario.

The existence of such systems of dwarfs is not unusual. Tully et al. (2006) pointed out that within 3~Mpc, except for KKR~25 and as we know KKs~3, every dwarf is associated with either a luminous group or an association of dwarfs. Makarov \& Uklein (2012) compiled a list of systems where the luminosity of both components are lower than the luminosity of the Small Magellanic Cloud. Such systems are quite numerous in the Local Supercluster and are located in low-density regions. In recent years, the number of discoveries of multiple dwarf systems have been made. For instance,  Makarova et al. (2018) discovered a satellite of an isolated irregular dwarf at a distance of 9.2~Mpc. This pair is characterized by a projected separation of 3.9~kpc and absolute $V$-magnitude of $-13.3$ and $-9.4$~mag for the main component and the satellite, respectively. Carlin et al. (2016) found a satellite with $M_g=-7.4$ at a projected distance of $\sim35$~kpc from dwarf spiral NGC~2403 in the framework of the MADCASH survey. This galaxy has an old metal-poor stellar population similar to ultra-faint galaxies of the Local Group. Chengalur \& Pustilnik (2013) discovered a triplet of extremely gas-rich galaxies near the center of the Lynx-Cancer void. The authors argued that the appearance of such objects in low-density regions is a consequence of the slow evolution of the structure formation in comparison with the denser regions.


The large uncertainty of its distance (see Sec. 3.1) and the lack of a radial velocity measurement prevent a clear association of Do~I with NGC 404. Another possibility is that Do~I could be an isolated dSph galaxy beyond the Local Group. Geha et al.(2002) found that quenched dwarf galaxies  are always found within 1.5~Mpc of a more massive host. However, while most of the known classical dSph galaxies have been
discovered within a distance of 300~kpc from the Milky Way or M31 (where environmental effects are important), there are already several known examples of presently isolated dwarf spheroidal galaxies.  In the case of a  distance for Do~I of 2.5~Mpc, it would be one of the most isolated dSph galaxies reported so far.

Fig.~8 shows a map in the super-galactic coordinates of the distribution of galaxies within a 5~Mpc cube roughly centered on the Local Group.  Known highly isolated dSph systems are marked with red stars, including  Do~I. 
The short line marked from Do~I indicates its position for the distance range from 2.5--3.5~Mpc discussed in Sec.~3.1. 
The list of isolated galaxies surrounding the Local Group is given in Table 3. All the isolated dSphs within 2 Mpc of the MW have precise distance measurements, which makes it possible to assert their isolation with great confidence (see column 7 in Table 3).
The dSph galaxies Cetus and And XVIII lie inside the zero-velocity surface of the Local Group, while Tucana is located just beyond the Local Group boundary at a distance of 0.91~Mpc from Milky Way and 1.38~Mpc from Andromeda (Jacobs et al. 2009). However, due to the
uncertain nature of their orbits, past interactions for these cannot
be excluded. Teyssier  et al. (2012) have analyzed the orbits of
potential dwarf galaxy halos in the Via Lactea~II simulation to
determine the likelihood that presently isolated dwarf galaxies had
past interactions with the Milky Way or M31. They found that some
$13\%$ of dwarf galaxies with present distances between 300 and
1500~kpc could be classified as escapees, making past interactions
likely. However, at a present distance larger than 2.5~Mpc, the origin of
Do~I as a past satellite of the Milky Way or M31 appears very unlikely,
which would make it the first unambiguous example of an isolated dwarf
spheroidal galaxy (if not a satellite of NGC~404). 
Apart from the gas fraction and the resultant star formation history,
which distinguish dwarf spheroidal and dwarf irregular galaxies (Grebel et al. 2003), past
tidal interactions can also be imprinted on a galaxy's stellar
kinematics (e.g., Mu\~noz, Majewski \& Johnston 2008).
 While not all dwarf spheroidals show such tidal features
(Pe\~narrubia et al. 2009), their presence in Do~I could be an
important further clue to its origin as a truly isolated dwarf
spheroidal, or as a system with past interactions of some kind.

\begin{table*}
\caption{Isolated dwarf galaxies around the Local Group \label{tbl-2}}
\tiny
\begin{tabular}{lrrrllclrr}
\hline\hline
    &         &  &  &  &  \multicolumn{2}{l}{\bf{Main Disturber}} & \multicolumn{3}{r}{\bf{Closest neighbor}} \\
Name    & \multicolumn{1}{c}{$M_B$} & \multicolumn{1}{c}{$M_V$} & \multicolumn{1}{c}{Distance} & Ref. &  & \multicolumn{1}{c}{$\Delta$} &   & \multicolumn{1}{c}{$M_B$}  & \multicolumn{1}{c}{$\Delta$} \\
    & \multicolumn{1}{c}{(mag)}   & \multicolumn{1}{c}{(mag)}   & \multicolumn{1}{c}{(Mpc)}      &                     &   & \multicolumn{1}{c}{(Mpc)}        &   & \multicolumn{1}{c}{(mag)}    & \multicolumn{1}{c}{(Mpc)}        \\
\hline
KKs~3     & $-10.8$ & $-12.3$ & $2.12\pm0.07$ & 1 & MW   & 2.12 & IC~3104       & $-14.9$ & 1.02 \\
KKR~25    & $ -9.4$ & $-10.7$ & $1.92\pm0.06$ & 2 & M~31 & 1.86 & KK~230        & $ -9.6$ & 0.98 \\
Tucana    & $ -9.2$ & $ -9.7$ & $0.92\pm0.02$ & 3 & MW   & 0.91 & ESO~245--007  & $ -9.5$ & 0.60 \\
Cetus     & $-10.1$ & $-10.7$ & $0.79\pm0.04$ & 3 & M~31 & 0.69 & IC~1613       & $-14.4$ & 0.22 \\
And XVIII &         & $-10.4$ & $1.33\pm0.08$ & 4 & M~31 & 0.55 & And~XIX       & $     $ & 0.44 \\
\hline\hline
  \multicolumn{10}{l}{References: (1)\cite{2015MNRAS.447L..85K}, (2)\cite{2012MNRAS.425..709M}  , (3)\cite {2009AJ....138..332J}, (4) \cite{2017MNRAS.464.2281M}}

\end{tabular}
\end{table*}

Because of its distance range and its crowding level, deeper, high resolution  {\it Hubble Space Telescope} imaging and radial velocity observations are needed to constrain the distance of Do~I and definitively determine if it is a NGC 404 companion.  The discovery of Do~I with an amateur telescope  also shows that ultra-deep imaging in wide sky areas is still very valuable for detecting faint,  hitherto unknown low surface brightness galaxies that were not detected by means of resolved stellar populations or H{\sc i} surveys. In fact, the wide fields and depths (2--3 magnitudes deeper than the POSS-II survey) of these small telescopes  make them ideal for uncovering galaxies within the Local Group and beyond ($\sim$ 10 Mpc) at low surface brightness levels, including in the outskirts of the Local Group where $\Lambda$-CDM simulations predict a significant population of free-floating, still undetected, isolated dwarf galaxies. These systems are of great interest for modern cosmological scenarios of galaxy formation, and could still remain undetected in large scale optical and radio surveys due to their extremely faint surface brightness and low gas content.

\begin{acknowledgements}


We acknowledge M. Cecconi and D. Carosati for their support during the follow-up of Do I with the TNG. We also thank N. Martin, C. Wittmann and G. Morales. DMD and EKG acknowledge support by Sonderforschungsbereich (SFB) 881
``The Milky Way System'' of the German Research Foundation (DFG), particularly through sub-project A2. TSC acknowledges the support of a National Science Foundation Graduate Research Fellowship. JAC-B acknowledges financial support from
CONICYT-Chile FONDECYT Postdoctoral Fellowship 3160502. MAB acknowledges support from grant AYA2016-77237-C3-1-P from the Spanish Ministry of Economy and Competitiveness (MINECO) and from the Severo Ochoa Excellence scheme (SEV-2015-0548). MPH is supported by grants from the US NSF/AST-1714828 and the
   Brinson Foundation. GD thanks Matteo Collina and Tim Stone for the use of their remote observatory for this work. 
   AJR was supported by National Science Foundation grant AST-1616710 and as a Research Corporation for Science Advancement Cottrell Scholar.
   We acknowledge the usage of the HyperLeda database (http://leda.univ-lyon1.fr). Based on observations made with the Italian Telescopio Nazionale
Galileo (TNG), operated by the Fundaci\'on Galileo Galilei of the INAF
(Istituto Nazionale di Astrofisica), and with the Gran Telescopio
Canarias (GTC) under Director's  Discretionary Time. Both telescopes are installed in the Spanish
Observatorio del Roque de los Muchachos (La Palma, Spain) of the
Instituto de Astrof\'\i sica de Canarias.

 The authors also acknowledge the use of IRAF, the Image Reduction and Analysis Facility, a general purpose software system for the reduction and analysis of astronomical data. IRAF is written and supported by the National Optical Astronomy Observatories (NOAO) in Tucson, Arizona. NOAO is operated by the Association of Universities for Research in Astronomy (AURA), Inc. under cooperative agreement with the National Science Foundation.
     
     Funding for SDSS-III has been provided by the Alfred P. Sloan Foundation, the Participating Institutions, the National Science Foundation, and the U.S. Department of Energy Office of Science. 

\end{acknowledgements}


\end{document}